\documentclass[aps,pra,twocolumn,showpacs,superscriptaddress]{revtex4}
\usepackage{graphicx,dcolumn,bm,amsmath,amssymb,amsfonts}
\begin{document}
\newcommand{\CNOT}{{\scriptsize CNOT} }
\newcommand{\vsigma}{\bm{\sigma}}
\newcommand{\tr}{\text{Tr}}
\title{Potential Errors in a Scheme of Universal Quantum Gates in Kane's Model}
\author{Yukihiro Ota}
 \email[Electronic address: ]{oota@hep.phys.waseda.ac.jp}
 \affiliation{Department of Physics, Waseda University, Tokyo 169--8555,
 Japan}
\author{Shuji Mikami}%
 \email[Electronic address: ]{mikami@hep.phys.waseda.ac.jp}
 \affiliation{Department of Physics, Waseda University, Tokyo 169--8555,
 Japan}
\author{Ichiro Ohba}
 \email[Electronic address: ]{ohba@waseda.jp}
 \affiliation{Department of Physics, Waseda University, Tokyo 169--8555,
 Japan}
 \affiliation{Kagami Memorial Laboratory for Material Science and
 Technology, Waseda University, Tokyo 169--0051, Japan}
 \affiliation{Advanced Research Center for Science and Technology,
 Waseda University, Tokyo 169--8555, Japan} 
\date{\today}
\begin{abstract}
We re--investigate a plausible proposal for universal quantum gates in
 Kane's model, in which the authors assumed that electron spin is 
 always downward under a background magnetic field and the value of
 controlling parameters is varied instantaneously. 
We demonstrate that a considerable error appears, for example, in the X
 rotation.  
As result, the controlled operations don't work. 
Such a failure is caused by improper choice of the computational bases;
 actually, the electron spin is not always downward over time during quantum operations. 
\end{abstract}
\maketitle
\section{Introduction}
Various physical models for quantum computation have been proposed. 
Among them, a nuclear magnetic resonance (NMR) quantum
computation by liquid state NMR\,\cite{Knill,Gershenfeld, Cory} is
the most successful, because several quantum algorithms
have experimentally worked, using five or seven qubits (see
Ref.\,\cite{Vandersypen} for additional references).  
On the other hand, Gershenfeld and Chuang\,\cite{Gershenfeld} pointed out that 
it was not possible to achieve beyond ten qubits using it.

Among many proposals for realizing quantum computation with
many qubits\,\cite{Kane, Ladd, Cory2000}, Kane\,\cite{Kane} proposed 
quantum computing by nuclear spins in a semiconductor.
In this proposal, the nuclear spin of a dopant atom, $^{31}{\rm P}$, implanted
into a silicon substrate is a single qubit, and quantum states are measured by
detecting currents of spin--polarized electrons around $^{31}{\rm P}$. 
Although the implantation of the atoms with requisite accuracy and the detection of the
single electronic charge motion are very difficult even in the present state of
the technology, Kane's proposal could achieve quantum computation by the
use of many qubits, applying the several existing microfabrication
techniques of semiconductors. 
Moreover, several important experimental techniques, for example, the single--ion
implantation method\,\cite{Shinada} and the single--electron
transistor\,\cite{Nakamura, Elzerman, Xiao}, have been developed
steadily.  
Hence, it is necessary to re--investigate the feasible conditions required
by Kane's original proposal, based on detailed theoretical analyses
and the present state of experimental techniques.  

Several investigations of the schemes of implementing quantum gates in
Kane's model have been reported.   
Wellard {\it et al.}\,\cite{Wellard} numerically derived an effective
Hamiltonian and proposed a nonadiabatic controlling scheme for a
controlled--{\scriptsize NOT} ({\scriptsize CNOT}) gate. 
It should be noted that the quantum computational bases chosen are not
eigenstates of their Hamiltonian.   
Fowler {\it et al.}\,\cite{Fowler} showed an adiabatic controlling
scheme for a \CNOT gate and studied the errors under the presence of
dephasing.  
Their proposal is based on Goan and Milburn's study, which is explained
briefly in Ref.\,\cite{Galindo}.  
Hill and Goan\,\cite{Hill} proposed a nonadiabatic scheme for arbitrary
single qubit operations and controlled operations (e.g., a \CNOT gate, a
swap gate, and a controlled--Z gate), and studied the effect of
dephasing.  

In particular, Hill and Goan's results are very important, as they make
it possible to construct a set of universal gates for quantum
computation\,\cite{Nielsen}.  
They use the assumption that spin of the donor electrons is always
downward, even if quantum gates are carried out. 
Furthermore, a nonadiabatic, or instantaneous, controlling process of
parameters is an ideal one from both theoretical and experimental points
of view.
Therefore, it is necessary to discuss the validity of such assumption and
idealization.

In this paper, we show that potential errors should exist in Hill and
Goan's schemes for the gates including spin--flip operations (e.g., X
rotations and \CNOT gates); there is a large difference between their
aimed state $|\phi_{target}\rangle$ obtained by quantum operations and
the physical state $|\phi_{phys}\rangle$ obtained by a more realistic
time evolution.  
We think that they didn't properly take account of the hyperfine
interaction (HY) between electron spins and nuclear spins in their
composition of the qubits system.

This paper is organized as follows.
First, we review Kane's model and, taking the strengths of all
interactions in the model into account, explain how to choose our
computational bases in Sec.\,\ref{sec:Model}. 
After the completion of choosing them, we discuss the schemes of
several important quantum gates, phase--shift operations, spin--flip
operations, and controlled--Z operations in Sec.\,\ref{sec:QG}. 
The main results are shown in Sec.\,\ref{subsec:spin_flip}. 
Finally, we summarize our results in Sec.\,\ref{sec:summary}. 
\section{Kane's model}\label{sec:Model}
Kane's model for two qubits is described by the Hamiltonian 
\begin{equation}
 H(t)=\sum_{i=1}^{2}H^{i}(t)+J(t)\vsigma^{1\,e}\cdot\vsigma^{2\,e}+\sum_{i=1}^{2}H_{ac}^{i}(t)\,,
\label{eq:2qubit_H}		
\end{equation}
where 
$H^{i}(t)=-g_{n}\mu_{n}B\sigma_{z}^{i\,n}+\mu_{B}B\sigma_{z}^{i\,e}+A_{i}(t)\vsigma^{i\,e}\cdot\vsigma^{i\,n}$,
$H^{i}_{ac}(t)=B_{ac}\bm{m}(t)\cdot(-g_{n}\mu_{n}\vsigma^{i\,n}+\mu_{B}\vsigma^{i\,e})$
and 
$\bm{m}(t)=(\cos(\omega_{ac}t),\,-\sin(\omega_{ac}t),\,0)$.
The Pauli matrix is  written by $\sigma_{k}$ $(k=x,\,y,\,z)$, and 
the superscript $ie$ ($in$) in it indicates the electronic (nuclear)
spin of the dopant atom implanted in the $i$ th site.  
According to Ref.\,\cite{Kane}, the value of the uniform static magnetic
field in the direction of the $z$ axis is $2.0\,{\rm T}$ (i.e.,
$B=2.0\,{\rm T}$). 
Then, the value of the Zeeman splitting energies for the electron,
$\mu_{B}B$, and the nucleus, $g_{n}\mu_{n}B$, are $0.116\,{\rm meV}$ and
$0.071\times 10^{-3}\,{\rm meV}$, respectively.   
The HY on the $i$ th site between the nuclear spin and the electronic spin, 
$A_{i}(t)\vsigma^{i\,e}\cdot\vsigma^{i\,n}$ is locally changed by
controlling the voltage in the $i$ th A--gate located right above the
$i$ th dopant atom\,\cite{Kane}.   
The typical value of the HY is $0.121\times 10^{-3}\,{\rm meV}$ 
($\equiv A_{0}$) when the voltage in the A--gate is vanishing, and the magnitude
of $A_{i}$ decreases as it increases\,\cite{Kane}. 
Notice that the nuclear Zeemann splitting energy and the
HY are almost the same order of magnitude. 
This causes a non--negligible mixing between the electron spin down and
up in the eigenstates of the Hamiltonian (see below). 
The electronic exchange interaction (EE) between the $1$st and the $2$nd
electronic spins, $J(t)\vsigma^{1\,e}\cdot\vsigma^{2\,e}$, 
is locally changed by controlling the voltage in the J--gate located between
the neighboring dopant atoms\,\cite{Kane}. 
We assume that it will be possible to change the magnitude of the EE
from $J=0$ to $J\simeq \mu_{B}B/2$\,\cite{Hill}.
The third term in Eq.\,(\ref{eq:2qubit_H}) is a time--dependent
interaction between the magnetic field perpendicular to the $z$ axis and
spins.
The typical value of $B_{ac}$ is assumed to be 
$B_{ac}\simeq 2.5\,{\rm mT}$\,\cite{Kane,Hill}.   
We write the eigenstates for $\sigma^{in}_{z}$ as $|0\rangle_{i}$ and
$|1\rangle_{i}$ ($\sigma^{in}_{z}|0\rangle_{i}=|0\rangle_{i}$ and
$\sigma^{in}_{z}|1\rangle_{i}=-|1\rangle_{i}$), and those for $\sigma^{ie}_{z}$
as $\lvert\uparrow\rangle_{i}$ and
$\lvert\downarrow\rangle_{i}$ ($\sigma^{ie}_{z}\lvert\uparrow\rangle_{i}=\lvert\uparrow\rangle_{i}$ and
$\sigma^{ie}_{z}\lvert\downarrow\rangle_{i}=-\lvert\downarrow\rangle_{i}$).  

To investigate the time evolution of the system in detail, we first
choose the states to represent a qubit from the eigenstates of
$H^{i}(t)$ with a fixed value of $A_{i}(t)$.
These eigenstates compose the temporal bases in the adiabatic calculation.
We abbreviate the argument $t$ in the equations of its eigenvalues and
eigenstates (we will recover it if necessary). 
Introducing $\epsilon=\mu_{B}B+g_{n}\mu_{n}B$, the eigenvalues
$E^{i}_{k}(A_{i})$ ($k=0,\,1,\,2,\,3$ from the below) are given as follows:  
\begin{eqnarray*}
E_{0}^{i}(A_{i})&=&-A_{i}-\sqrt{\epsilon^{2}+4A_{i}^{2}}, \\ 
E_{1}^{i}(A_{i})&=& -\mu_{B}B+g_{n}\mu_{n}B+A_{i}, \\
E_{2}^{i}(A_{i})&=& -A_{i}+\sqrt{\epsilon^{2}+4A_{i}^{2}}, \\
E_{3}^{i}(A_{i})&=& \mu_{B}B-g_{n}\mu_{n}B+A_{i}. 
\end{eqnarray*}
Defining $\cos\theta_{i}=(\epsilon+\sqrt{\epsilon^{2}+4A_{i}^{2}})/N_{i}$, 
$\sin\theta_{i}=2A_{i}/N_{i}$, and 
$N_{i}^{2}=2\sqrt{\epsilon^{2}+4A_{i}^{2}}(\epsilon+\sqrt{\epsilon^{2}+4A_{i}^{2}})$, 
their corresponding eigenstates $\lvert u_{k}(A_{i})\rangle_{i}$ are
also given by 
\begin{eqnarray*}
\lvert u_{0}(A_{i})\rangle_{i}&=&
-\sin\theta_{i}\,\lvert\uparrow\rangle_{i}|1\rangle_{i}
+\cos\theta_{i}\,\lvert\downarrow\rangle_{i}|0\rangle_{i}, \\ 
\lvert u_{1}(A_{i})\rangle_{i}&=&\lvert\downarrow\rangle_{i}|1\rangle_{i}, \\
\lvert u_{2}(A_{i})\rangle_{i}&=&
\cos\theta_{i}\,\lvert\uparrow\rangle_{i}|1\rangle_{i}
+\sin\theta_{i}\,\lvert\downarrow\rangle_{i}|0\rangle_{i}, \\
\lvert
 u_{3}(A_{i})\rangle_{i}&=&\lvert\uparrow\rangle_{i}|0\rangle_{i}. 
\end{eqnarray*}
The operator 
$S_{z}^{i}\equiv(\sigma_{z}^{ie}+\sigma_{z}^{in})/2$ commutes with
$H^{i}$: 
$S_{z}^{i}\lvert u_{0}(A_{i})\rangle_{i}=S_{z}^{i}\lvert u_{2}(A_{i})\rangle_{i}=0$,
$S_{z}^{i}\lvert u_{1}(A_{i})\rangle_{i}=-\lvert u_{1}(A_{i})\rangle_{i}$,
and 
$S_{z}^{i}\lvert u_{3}(A_{i})\rangle_{i}=\lvert u_{3}(A_{i})\rangle_{i}$.
The ground state is always $\lvert u_{0}(A_{i})\rangle_{i}$. 
The energy differences $\Delta_{k0}^{i}\equiv E^{i}_{k}(A_{i})-E^{i}_{0}(A_{i})$
between the $k$ th excited states and the ground state are estimated as follows: 
$\Delta_{10}^{i} \simeq 2A_{i}+2g_{n}\mu_{n}B$, 
$\Delta_{20}^{i} \simeq 2\epsilon$, and 
$\Delta_{30}^{i} \simeq 2\epsilon +2A_{i}-2g_{n}\mu_{n}B$.
The values of two terms in $\Delta^{i}_{10}$ are of the same order of
magnitude as HY; it is possible to control the phase difference between 
$\lvert u_{0}(A_{i})\rangle_{i}$ and $\lvert u_{1}(A_{i})\rangle_{i}$, varying the
magnitude of the HY, $A_{i}(t)$. 
Note that these two states are robust against this controlling 
process because they are the eigenstates of temporal Hamiltonian
$H^{i}(t)$. 
Consequently, we choose the quantum computational bases for a single qubit
as $\lvert u_{0}(A_{i})\rangle_{i}$ and $\lvert u_{1}(A_{i})\rangle_{i}$. 
In particular, we write $|0\rangle_{L,i}\equiv\lvert u_{0}(A_{0})\rangle_{i}$ and 
$|1\rangle_{L,i}\equiv\lvert u_{1}(A_{0})\rangle_{i}$ when $A_{i}(t)=A_{0}$. 

Next, we discuss polarization of the electron spins.
Taking account of $\sin\theta_{i}\simeq 10^{-3}$ for $A_{i}=A_{0}$, we
may regard $\lvert u_{0}(A_{i})\rangle_{i}\simeq\lvert\downarrow\rangle_{i}|0\rangle_{i}$,
and the electrons are almost polarized in the downward direction of the $z$ axis.
If the system is a thermal equilibrium state with the 
temperature $T=100\,{\rm mK}$ 
($\beta^{-1}=k_{B}T\simeq 8.62\times 10^{-6}\,{\rm eV}$), 
which is a typical value in Kane's model\,\cite{Kane}, 
the ratio of the number of upper polarized electrons in the
$i$ th site, $n^{ie}_{\uparrow}$, to the number of downward polarized
ones in it, $n^{ie}_{\downarrow}$, is estimated as follows:
\begin{eqnarray*}
\frac{n^{ie}_{\uparrow}}{n^{ie}_{\downarrow}}
&\equiv&
\frac{\tr(\lvert\uparrow\rangle_{i}\langle\uparrow\rvert e^{-\beta
H^{i}})}{\tr(\lvert\downarrow\rangle_{i}\langle\downarrow\rvert
e^{-\beta H^{i}})} \\
&=&
\frac{e^{-\beta\Delta^{i}_{30}}+\cos^{2}\theta_{i}\,e^{-\beta\Delta_{20}^{i}}+\sin^{2}\theta_{i}}{\sin^{2}\theta_{i}\,e^{-\beta\Delta^{i}_{20}}+e^{-\beta\Delta_{10}^{i}}+\cos^{2}\theta_{i}}
\\
&\simeq& 10^{-5}.
\end{eqnarray*}
Here, we have used the following estimation: $\sin\theta_{i}\simeq 10^{-3}$, 
$2\beta\epsilon\simeq 2\mu_{B}B/k_{B}T\simeq 10$, and 
$2\beta A_{0}\simeq 10^{-2}$. 
The result, $n^{ie}_{\uparrow}/n^{ie}_{\downarrow}\simeq 10^{-5}$, 
also suggests that the electron spins should be $\lvert\downarrow\rangle_{i}$.  
However, the above consideration is based on the static physical
property; it is nontrivial that it could sustain the downward
polarization through the temporal controlling process.

Now, we address the preparation of an initial state. 
The thermal equilibrium state of the system 
$\rho_{eq}=\otimes_{i} (e^{-\beta H^{i}}/\tr\,e^{-\beta H^{i}})$ 
is not a pure state, even if the temperature $T$ is $100\,{\rm mK}$. 
In fact, we find  
$\rho_{eq}^{i}\simeq 0.508 \,\lvert u_{0}(A_{i})\rangle_{i}\langle u_{0}(A_{i})\rvert+0.492 \,\lvert u_{1}(A_{i})\rangle_{i}\langle u_{1}(A_{i})\rvert$ 
for $T=100\,{\rm mK}$, $A_{i}=A_{0}$, $J=0$, and $B=2.0\,{\rm T}$. 
The initialization problem might be solved by the standard scheme in
liquid NMR (e.g., by the use of effective pure states). 
Throughout this paper, we investigate only whether a given pure state evolves to a desired
state by controlling processes.
Such consideration is meaningful, because any arbitrary mixed
state is represented by suitable linear combinations of pure states,
and the control of a quantum system according to our expectations will
be necessary to overcome initialization problems.  

\begin{figure}[htbp]
\centering
\scalebox{0.50}[0.50]{\includegraphics{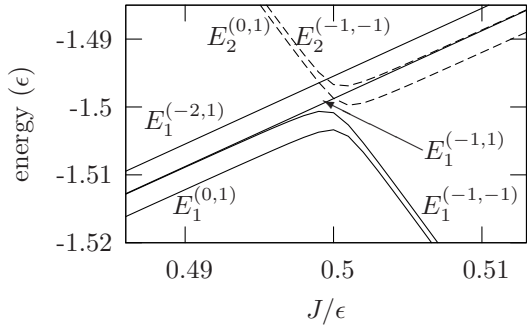}}
\caption{Configurations of the $H_{C}$ energy level around the level
 crossing point ($J\simeq \mu_{B}B/2$). Four solid lines represent the lowest
 eigenvalues of $H_{C}^{(0,\,1)}$, $H_{C}^{(-1,\,-1)}$,
 $H_{C}^{(-1,\,1)}$, and $H_{C}^{(-2,\,1)}$, respectively, which are
 associated with the computational bases. Two dashed lines represent the second
 lowest eigenvalues for each $H_{C}^{(0,\,1)}$ and $H_{C}^{(-1,\,1)}$.} 
\label{fig:c_elv}
\end{figure}
Finally, let us consider the eigenvalue problem of the two qubit 
Hamiltonian
\begin{equation*}
H_{C}(t)\equiv\sum_{i=1}^{2}H^{i}(t)+J(t)\vsigma^{1\,e}\cdot\vsigma^{2\,e}, 
\end{equation*}
fixing the values of $A_{1}(t)$, $A_{2}(t)$, and $J(t)$. 
The Hamiltonian $H_{C}(t)$ is  composed of the first and the second terms
in Eq.\,(\ref{eq:2qubit_H}) and plays a central role in the controlled
operations.  
Hereafter, we write $\lvert\psi\rangle_{i}\lvert\phi\rangle_{i}$ as
$\lvert\psi\phi\rangle_{i}$ (e.g.,
$\lvert\downarrow\rangle_{i}|1\rangle_{i}=\lvert\downarrow 1\rangle_{i}$). 
We introduce the total spin operator in the direction of the $z$ axis
$S^{tot}_{z}=(\sigma_{z}^{1\,n}+\sigma_{z}^{1\,e}+\sigma_{z}^{2\,n}+\sigma_{z}^{2\,e})/2$ 
and the parity operator, $P$ defined as follows:
$P^{\dagger}\sigma_{a}^{1\alpha}P=\sigma_{a}^{2\alpha}$ and 
$P^{\dagger}\sigma_{a}^{2\alpha}P=\sigma_{a}^{1\alpha}$ ($\alpha=e,\,n$
and $a=x,\,y,\,z$).  
The operator $P$ is an idempotent operator: $P^{2}=\openone$. 
We find that $S^{tot}_{z}$ always commutes with both $H_{C}(t)$ and $P$,
and $P$ commutes with $H_{C}(t)$ if $A_{1}(t)=A_{2}(t)$. 
As result, in the case of $A_{1}(t)=A_{2}(t)$, the Hamiltonian $H_{C}(t)$ becomes a block
diagonal matrix whose blocks are characterized by the eigenvalues of
$S^{tot}_{z}$ and $P$,
$s$ $(\in\sigma(S^{tot}_{z})\equiv\{2,\,1,\,0,\,-1,\,-2\})$ and
$p$ $(\in\sigma(P)\equiv\{1,\,-1\})$, respectively: 
\begin{equation}
H_{C}(t)=\bigoplus_{s,\,p}H_{C}^{(s,\,p)}(t).
\label{eq:c_h_dcmp}
\end{equation}
Note that
$S^{tot}_{z}H^{(s,\,p)}_{C}(t)=H^{(s,\,p)}_{C}(t)S^{tot}_{z}=sH^{(s,\,p)}_{C}(t)$ and
$PH^{(s,\,p)}_{C}(t)=H^{(s,\,p)}_{C}(t)P=pH^{(s,\,p)}_{C}(t)$.
A similar result is given by Berman\,{\it et al.}\,\cite{Berman}, but
our computational bases are different from theirs.
The submatrices $H_{C}^{(s,p)}(t)$ in Eq.\,(\ref{eq:c_h_dcmp}) are
$4\times 4$ ones at most. 
Hence, we calculate analytically the diagonal form of Eq.\,(\ref{eq:c_h_dcmp}).
We show the explicit forms of its eigenvalues and eigenstates in 
Appendix\,\ref{append:ana_eq}. 
We explain the relationship between the eigenstates of
$H_{C}(t)$ and the computational bases.
If the value of $J(t)$ is vanishing, we find that 
$\lvert v_{1}\rangle\equiv |0\rangle_{L,1}|0\rangle_{L,2}$, 
$\lvert v_{+}\rangle \equiv (|0\rangle_{L,1}|1\rangle_{L,2} + |1\rangle_{L,1}|0\rangle_{L,2})/\sqrt{2}$,  
$\lvert v_{-}\rangle \equiv (|0\rangle_{L,1}|1\rangle_{L,2} - |1\rangle_{L,1}|0\rangle_{L,2})/\sqrt{2}$,
and $\lvert v_{4}\rangle\equiv|1\rangle_{L,1}|1\rangle_{L,2}$ are the eigenstates of
$H^{(0,1)}_{C}(t)$, $H_{C}^{(-1,1)}(t)$, $H_{C}^{(-1,-1)}(t)$, and
$H_{C}^{(-2,+1)}(t)$, respectively. 
Therefore, these four states play an essential role in the scheme of
controlled operations. 
In Fig.\,\ref{fig:c_elv}, we show that the eigenvalues of submatrices
related to the computational bases around the level crossing
($J\simeq\mu_{B}B/2$), varying the magnitude of the EE, $J$.
The eigenvalues of $H_{C}^{(s,\,p)}$ are specified by $E^{(s,\,p)}_{k}$
($k=1,\,2$ for $(s,\,p)=(0,\,1),\,(-1,\,1)$, and $k=1$ for
$(s,\,p)=(-1,\,1),\,(-2,\,1)$). 
The energy difference $E_{1}^{(-1,\,1)}-E_{1}^{(-1,\,-1)}$ is tiny and
positive for $J<\mu_{B}B/2$. 
As $J$ increases, the levels $E_{1}^{(0,\,1)}$ and $E_{1}^{(-1,\,-1)}$
approach $E_{2}^{(0,\,1)}$ and $E_{2}^{(-1,\,-1)}$, respectively. 
\section{The quantum gates}\label{sec:QG}
\subsection{The phase--shift operations}\label{subsec:PS}
We achieve the phase--shift operation for the $i$ th qubit, varying
$A_{i}(t)$ adiabatically with the other parameters fixed; $J(t)=0$,
$B_{ac}=0$, and $A_{j}(t)=A_{0}$ ($j\neq i$).

\begin{figure}[htbp]
\centering
\scalebox{0.5}[0.5]{\includegraphics{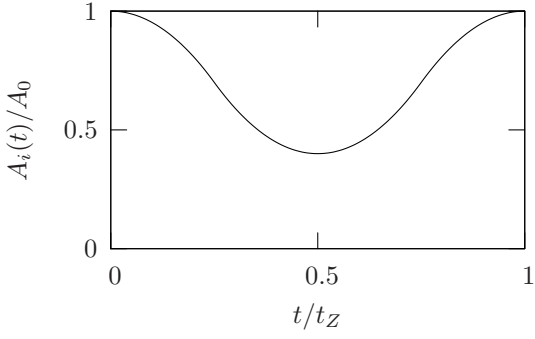}}
\caption{Profile of $A_{i}(t)$ ($a=0.6$).} 
\label{fig:profile_A}
\end{figure}

First, we calculate the time--evolution operator for the $i$ th qubit: 
\begin{equation}
U^{i}(t_{Z})
=
T\left\{\exp\left[-\frac{i}{\hbar}\int_{0}^{t_{Z}}H^{i}(t)\,dt\right]\right\}. 
\label{eq:s_t_evl}
\end{equation}
The symbol $T$ means the time--ordered product. 
We define the operation time for the phase--shift operation as $t_{Z}(>0)$.
Introducing a dimensionless parameter $a$ ($0<a<1$) and a dimensionless
variable $\tau=t/t_{Z}$, we assume the time dependence of $A_{i}(t)$ as
follows (see, Fig.\,\ref{fig:profile_A}):  
\begin{equation}
A_{i}(t)=\left\{
\begin{array}{ll}
A_{0}f_{A}(\tau) & \left(0\le \tau \le \frac{1}{2}\right)\\
A_{0}f_{A}(1-\tau) & \left(\frac{1}{2}\le \tau \le 1 \right)
\end{array} 
\right. .\label{eq:HY_profile}
\end{equation}

The function $f(\tau)$ in Eq.\,(\ref{eq:HY_profile}) is defined as 
\begin{eqnarray*}
f_{A}(\tau)
&=&
\Theta\left(\frac{1}{4}-\tau\right)\left(1-8a\tau^{2}\right) \\
&&
\quad\quad
+
\Theta\left(\tau-\frac{1}{4}\right)
\left[1-a+8a\left(\tau-\frac{1}{2}\right)^{2}\right], 
\end{eqnarray*}
and $\Theta(\tau)$ is the Heaviside function. 
The initial state for the $i$ th qubit will be spanned by $|0\rangle_{L,i}$ and
$|1\rangle_{L,i}$, because $A_{i}(0)=A_{0}$. 
We also find that $\lvert u_{0}(A_{i}(t_{Z}))\rangle_{i}=|0\rangle_{L,i}$ and
$\lvert u_{1}(A_{i}(t_{Z}))\rangle_{i}=|1\rangle_{L,i}$, because we assume
that the time dependence of $A_{i}(t)$ is given by
Eq.\,(\ref{eq:HY_profile}).
Consequently, Eq.\,(\ref{eq:s_t_evl}), due to the adiabatic
theorem\,\cite{Kato, Messiah}, is written by    
\begin{equation}
U^{i}(t_{Z})=e^{-i\delta^{i}_{0}}|0\rangle_{L,i}\langle
 0|+e^{-i\delta^{i}_{1}}|1\rangle_{L,i}\langle 1| +R^{i},  
\label{eq:s_ad_t_ev}
\end{equation}
where
$\delta^{i}_{0}=t_{Z}h^{-1}\int_{0}^{1}\left[-A_{i}(s)-\sqrt{\epsilon^{2}+4A_{i}(s)^{2}}\right]ds$
and 
$\delta^{i}_{1}
=
t_{Z}\hbar^{-1}\left[g_{n}\mu_{n}B-\mu_{B}B+\int_{0}^{1}A_{i}(s)\,ds\right]$.
The third term in Eq.\,(\ref{eq:s_ad_t_ev}), $R^{i}$ includes 
$\lvert u_{2}(A_{i}(t_{Z}))\rangle_{i}\langle u_{0}(A_{i}(t_{Z}))|$ and 
$\lvert u_{0}(A_{i}(t_{Z}))\rangle_{i}\langle u_{2}(A_{i}(t_{Z}))|$
mediated by the level crossing effect in the course of operation.
By the definition of a matrix norm\,\cite{Horn} for an $n\times n$ complex
matrix $M=(M_{ab})$, 
$\|M\|\equiv \max_{1\le a,b\le n}|M_{ab}|$, 
$\|R^{i}\|$ is estimated as the order of $\hbar/2\epsilon t_{Z}$, and
later on we find that it's very small. 
A criterion for the validity of the adiabatic approximation is
given as follows\,\cite{Messiah}: 
\begin{equation}
\max_{0\le\tau\le 1}\left|\frac{
\langle u_{2}^{i}(\tau)
|d H^{i}(\tau) /d\tau |u_{0}^{i}(\tau)\rangle
}{E_{2}^{i}(\tau)-E_{0}^{i}(\tau)}
\right|<1. \label{eq:s_ad_cond}
\end{equation}
In this case, we have a condition,   
\begin{equation*}
\frac{\epsilon A_{0}}{\epsilon^{2}+4A_{0}^{2}}a <1. 
\end{equation*} 
The inequality\,(\ref{eq:s_ad_cond}) is always fulfilled,
because $O(2A_{0} /\epsilon)\sim 10^{-3}$ and $O(a) \sim 1$. 

Next, we calculate the time--evolution operator for the remaining $j$ th
qubit ($j\neq i$).  
Taking account of the constancy of $A_{j}(t)=A_{0}$ and the initial
states for the $j$ th qubit spanned by $|0\rangle_{L,j}$ and
$|1\rangle_{L,j}$, we simply obtain the
following result:   
\begin{equation}
U^{j}(t_{Z})
=
e^{-i\delta^{j}_{0}}|0\rangle_{L,j}\langle 0|
+e^{-i\delta^{j}_{1}}|1\rangle_{L,j}\langle 1|,
\label{eq:s_other_ev}
\end{equation}
where 
$\delta^{j}_{0} =t_{Z}\hbar^{-1}(-A_{0}-\sqrt{\epsilon^{2}+4A_{0}^{2}})$
and 
$\delta^{j}_{1} =t_{Z}\hbar^{-1}(g_{n}\mu_{n}B-\mu_{B}B+A_{0})$. 

Now, we construct the phase--shift operation (i.e., a Z rotation) with
the given angle $\theta_{Z}$ for the $i$ th qubit. 
In the first place, the phase difference between $|0\rangle_{L,i}$ and
$|1\rangle_{L,i}$ should be equal to $\theta_{Z}$; 
$\delta^{i}_{0}-\delta^{i}_{1}=\theta_{Z}+2n \pi$ ($n\in\mathbb{Z}$), due to
Eq.\,(\ref{eq:s_ad_t_ev}).
Secondly, for all qubits except for  the $i$ th qubit, no phase differences have
to exist; $\delta^{j}_{0}-\delta^{j}_{1}=2m \pi $ ($m\in\mathbb{Z}$,
$m\neq 0$, and $j\neq i$), due to Eq.\,(\ref{eq:s_other_ev}). 
Using these two conditions, we obtain the following equations:
\begin{widetext}
\begin{eqnarray}
&&
\frac{\epsilon-2g_{n}\mu_{n}B+\int_{0}^{1}\left[
-2A_{i}(s)-\sqrt{\epsilon^{2}+4A_{i}(s)^{2}}\,ds
\right]}{\epsilon-2g_{n}\mu_{n}B-2A_{0}-\sqrt{\epsilon^{2}+4A_{0}^{2}}}
=
\frac{\theta_{Z}+2n \pi}{2m \pi}, \label{eq:s_ph_cond1}\\
&&
t_{Z}
=
\frac{2m \pi \hbar}{
\epsilon-2g_{n}\mu_{n}B-2A_{0}-\sqrt{\epsilon^{2}+4A_{0}^{2}}}\label{eq:s_ph_cond2}. 
\end{eqnarray} 
\end{widetext}
We find that Eq.\,(\ref{eq:s_ph_cond1}) includes only one unknown
parameter $a$, except for the free parameters $m$ and $n$. 
We should choose suitable values for them so that the solution of
$O(a)\sim 1$ exists for any $\theta_{Z}$; actually, we choose $m=-5$ and $n=-6$. 
We numerically solve Eq.\,(\ref{eq:s_ph_cond1}) by the bisection method.   
Then, we get the operation time $t_{Z}$ by substituting the value of
$m$ for Eq.\,(\ref{eq:s_ph_cond2}); its value is independent of the
value of $\theta_{Z}$. 

We show the results in Table\,\ref{tab:s_ph_result}.
The minimum value of $A_{i}(t)$ is given by $A_{0}(1-a)$. 
The phase--shift operations of the angle $\theta_{Z}=\pi/4$
(i.e., the $\pi/8$ gate) and $\pi/2$ (i.e., the phase gate) are
performed by reducing the HY by up to about $66$\% during the operation
time, which is about two times longer than that of Ref.\,\cite{Hill}. 
This difference comes from our chosen profile function of $A_{i}(t)$, valid
for the adiabatic approximation.  
The value of $\hbar/2\epsilon t_{Z}$ is approximately estimated as
$A_{0}/4 |m| \pi \epsilon$; $A_{0}/4 |m| \pi \epsilon\simeq 10^{-5}$ for
$m=-5$; the correction to the adiabatic approximation is very small. 
The choice of $m$ and $n$ is not unique. 
In fact, we find that other solutions of Eq.\,(\ref{eq:s_ph_cond1}) exist
if $m\le -5$ and $n\le -6$. 
The operation time $t_{Z}$ for $(m,\,n)=(-5,\,-6)$ is the shortest of the
solutions.  
\begin{table}
\caption{\label{tab:s_ph_result} Calculated values of $a$ and $t_{Z}$ with $m=-5$ and $n=-6$. }
\begin{ruledtabular}
\begin{tabular}{ccc}
$\theta_{Z}$ & $a$ & $t_{Z}$ ($\mu s$)\\
\hline
$\pi/4$ & 0.598 & 0.05 \\
$\pi/2$ & 0.664 & 0.05
\end{tabular}
\end{ruledtabular}
\end{table}
We confirm also the realization of the desired phase--shift gates
by solving numerically the time--dependent Schr\"odinger equation for
the resultant parameters.

We summarize the results for the phase--shift operation. 
We have achieved it by adiabatically controlling the magnitude of the
HY, $A_{i}(t)$, with $J(t)=0$, $A_{j}(t)=A_{0}$ ($j\neq i$), and
$B_{ac}=0$. 
Its errors are very small, as the adiabatic approximation works well. 
The phase--shift operations relevant to universal quantum gates, the $\pi/8$ gate
($\theta_{Z}=\pi/4$) and the phase gate ($\theta_{Z}=\pi/2$), are
realized by reducing the value of $A_{i}(t)$ from $A_{0}$ to about
$0.4 A_{0}$ at most during the operation time $t_{Z}(\simeq 0.05\,\mu s)$. 
According to \cite{Kane}, it might be possible to decrease the magnitude of
the HY by up to $50$\%. 
This means that further development of the experimental techniques might
 be needed to fully realize them by our scheme. 
\subsection{The spin--flip operations}\label{subsec:spin_flip}
Let us consider the scheme of the spin--flip operation. 
We reproduce the spin--flip operation in Ref.\,\cite{Hill} by applying the transverse
time--dependent magnetic field (see the third term in
Eq.\,(\ref{eq:2qubit_H})), using an approximation. 
The idea is based on a standard technique in NMR\,\cite{Abragam,Slichter}.  
Here, we assume that the transverse magnetic field is turned on or off
instantaneously. 
We show that a serious error exists in Ref.\,\cite{Hill}, even if such
an ideal controlling process is realized.   
Hereafter, we concentrate on the scheme of an X rotation for the $i$ th
qubit.

First of all, we explain the controlling processes of the EE, the HY,
and the transverse time--dependent magnetic field.    
We keep $J(t)=0$ and $A_{j}(t)=A_{0}$ ($j\neq i$). 
By contrast, we adiabatically change the magnitude of $A_{i}(t)$; 
in the first place, we decrease the value of $A_{i}(t)$ up to a suitable
constant value $A(<A_{0})$ (we explain how to choose the value of $A$
below) during the time interval $t^{\prime}_{X}/2$ (the first step), keep $A_{i}(t)=A$ during 
the time interval $t_{X}$ (the second step), and finally increase it from $A$ to
$A_{0}$ (the third step). 
Thus, the time dependence of $A_{i}(t)$ is given by 
\begin{equation}
A_{i}(t)=\left\{
\begin{array}{ll}
A_{0}f_{A}(\tau) & \left(0\le \tau \le \frac{1}{2}\right)\\
A & \left(\frac{1}{2}\le\tau\le\tau_{sp}-\frac{1}{2}\right)\\
A_{0}f_{A}(\tau_{sp}-\tau) & \left(\tau_{sp}-\frac{1}{2}\le \tau \le \tau_{sp} \right)
\end{array} 
\right. ,\label{eq:HY_profile_sp}
\end{equation}
where $\tau=t/t^{\prime}_{X}$, $\tau_{X}=t_{X}/t^{\prime}_{X}$, and 
$\tau_{sp}=1+\tau_{X}$. 
The parameter $a$ in $f_{A}(\tau)$ is given by $a=1-A/A_{0}$. 
In addition, the transverse time--dependent magnetic field is globally
applied, while we keep $A_{i}(t)=A$ (the second step). 

A undesired phase difference between $|0\rangle_{i}$ and $|1\rangle_{i}$
is caused by the adiabatically varying processes during $A_{i}(t)$ (the first
and third steps). 
This phase difference can be canceled by carrying out a suitable Z
rotation, as described in Sec.\,\ref{subsec:PS}, before and after the
first and third steps respectively.

Let us discuss the Hamiltonian in the second step.  
After instantaneously  turning on the transverse magnetic field, the
Schr\"odinger equation for the $i$ th qubit is written from
Eq.\,(\ref{eq:2qubit_H}):  
\begin{equation}
i\hbar\frac{d}{dt}\lvert\psi(t)\rangle_{i}
= 
\left[
H^{i} +H^{i}_{ac}(t)
\right]\lvert\psi(t)\rangle_{i}.  
\label{eq:flip_lab_eq}
\end{equation}
We solve Eq.\,(\ref{eq:flip_lab_eq}) in the rotating
frame\,\cite{Abragam,Slichter}; we remove the time dependence of the
Hamiltonian in Eq.\,(\ref{eq:flip_lab_eq}) by applying the unitary
operator 
$D_{z}(t;\omega_{ac})\equiv\exp\left[-i\,\omega_{ac} t(\sigma_{z}^{i\,e}+\sigma_{z}^{i\,n}) /2\right]$.
Introducing $\lvert\psi_{rot}(t)\rangle_{i}\equiv D_{z}(t;\omega_{ac})\lvert\psi(t)\rangle_{i}$, 
we obtain the following Schr\"odinger equation in the rotating frame
from Eq.\,(\ref{eq:flip_lab_eq}):  
\begin{equation}
i\hbar\frac{d}{dt}\lvert\psi_{rot}(t)\rangle_{i} 
=
H^{i}_{rot} \lvert\psi_{rot}(t)\rangle_{i}, \label{eq:flip_rot_eq} 
\end{equation}
where $H^{i}_{rot} = \hbar\omega^{e}\sigma_{z}^{i\,e}/2-\hbar\omega^{n}\sigma_{z}^{i\,n}/2+A\vsigma^{i\,e}\cdot\vsigma^{i\,n}+\mu_{B}B_{ac}\sigma_{x}^{i\,e}-g_{n}\mu_{n}B_{ac}\sigma_{x}^{i\,n}$,
$\hbar\omega^{e}=2\mu_{B}B+\hbar\omega_{ac}$, and
$\hbar\omega^{n}=2g_{n}\mu_{n}B-\hbar\omega_{ac}$. 
We naturally expand $\lvert\psi_{rot}(t)\rangle_{i}$ and $H^{i}_{rot}$ by 
$\{\lvert u_{k}(A)\rangle_{i}\}$. 
Let us introduce the matrix representation for an operator $M$ as follows: 
$M_{ab}\equiv\,_{i}\langle u_{a}(A)\lvert M\rvert u_{b}(A)\rangle_{i}$ 
 ($a,\,b=0,\,1,\,2,\,3$).
Thus, the Hamiltonian in the rotating frame, $H^{i}_{rot}$, is written
down as the sum of a block diagonal matrix $H^{i}_{d}$ and a off--diagonal
matrix $H_{mix}^{i}$: 
\begin{widetext}
\begin{eqnarray}
H^{i}_{rot} &=&
H_{d}^{i}+H_{mix} \nonumber \\
&=&
\left(
\begin{array}{cccc}
E^{i}_{0}(A) & -\nu_{\theta}B_{ac} & 0& 0\\
-\nu_{\theta}B_{ac} & E^{i}_{1}(A)-\hbar\omega_{ac} & 0& 0\\
0 & 0& E_{2}^{i}(A)& -\nu_{-\theta}B_{ac} \\
0 & 0& -\nu_{-\theta}B_{ac} & E^{i}_{3}(A)+\hbar\omega_{ac} 
\end{array}
\right) 
+ \left(
\begin{array}{cccc}
0 & 0& 0& \mu_{-\theta}B_{ac}\\
0 & 0& \mu_{\theta}B_{ac}& 0 \\
0 & \mu_{\theta}B_{ac}& 0& 0\\
\mu_{-\theta}B_{ac} & 0& 0& 0
\end{array}
\right).\label{eq:sf_H_mrep}
\end{eqnarray}
\end{widetext}
Here, we define $\mu_{\theta}=\mu_{B}\cos\theta-g_{n}\mu_{n}\sin\theta$ and 
$\nu_{\theta}=\mu_{B}\sin\theta+g_{n}\mu_{n}\cos\theta$, and the mixing
angle $\theta$ is defined for $\theta_{i}$ when $A_{i}(t)=A$. 
Whenever one can safely assume that the off--diagonal Hamiltonian,
$H^{i}_{mix}$, is negligible, the computational bases, 
$\lvert u_{0}(A)\rangle_{i}$ and $\lvert u_{1}(A)\rangle_{i}$, do not mix
with the  remaining irrelevant states, $\lvert u_{2}(A)\rangle_{i}$ and
$\lvert u_{3}(A)\rangle_{i}$, because of the block diagonal Hamiltonian,
$H^{i}_{d}$. 
In most studies up to now, this approximation has been taken for granted. 

However, it is nontrivial whether we should omit $H^{i}_{mix}$ in solving
Eq.\,(\ref{eq:flip_rot_eq}).
Notice that $\mu_{\theta}$ is larger than $\nu_{\theta}$, unless
$\theta$ is close to $\pi/2$.
We show that a considerable error is caused by neglecting
$H^{i}_{mix}$ below.

Before discussing such an error, we reproduce the results in
Ref.\,\cite{Hill} by means of our computational bases.
We solve Eq.\,(\ref{eq:flip_rot_eq}), omitting $H^{i}_{mix}$. 
We choose the value of $A$ so as to satisfy the Larmor resonance
condition corresponding to the computational bases, 
\begin{equation}
\hbar\omega_{ac}=E^{i}_{1}(A)-E_{0}^{i}(A)\label{eq:Larmor}.
\end{equation}
Taking account of Eq.\,(\ref{eq:Larmor}) and the initial state
$\lvert\phi\rangle_{i}$ expanded by $\lvert u_{0}(A)\rangle_{i}$ and 
$\lvert u_{1}(A)\rangle_{i}$ (i.e., the state just before turning on the
transverse magnetic field), 
we obtain the resultant state at $t_{X}$ in the rotating frame: 
\begin{equation}
\lvert\psi_{rot}(t_{X})\rangle_{i}
=
e^{-\frac{i}{\hbar}t_{X}E_{0}^{i}(A)}\,e^{i\,\theta_{X}\,\sigma^{i}_{X}(A)}\lvert\phi\rangle_{i}.
\label{eq:rot_sol_approx}
\end{equation}
We define 
$\sigma_{X}^{i}(A)=\lvert u_{0}(A)\rangle_{i}\langle u_{1}(A)\rvert + \lvert u_{1}(A)\rangle_{i}\langle u_{0}(A)\rvert$ 
and $\theta_{X}=\nu_{\theta}B_{ac}t_{X}/\hbar$.
Accordingly, 
\begin{equation}
\lvert\psi(t_{X})\rangle_{i}
=
e^{-\frac{i}{\hbar}t_{X}E_{0}^{i}(A)}\,
D^{\dagger}_{z}(\omega_{ac}t_{X})\,
e^{i\,\theta_{X}\,\sigma^{i}_{X}(A)}\lvert\phi\rangle_{i}. 
\label{eq:vec_lab_approx}
\end{equation}
Under the Larmor resonance condition (\ref{eq:Larmor}) for the $i$ th
qubit, the time--evolution of the remaining $j$ th qubit ($j\neq i$) is
approximately expressed by a Z rotation.
Moreover, both the overall phase and $D^{\dagger}_{z}(\omega_{ac}t_{X})$ in
Eq.(\ref{eq:vec_lab_approx}) are canceled by carrying out suitable Z
rotations. 
Consequently, we make only the state of the target qubit flip due to
Eq.\,(\ref{eq:vec_lab_approx}). 
In this way, we have reproduced a result corresponding to that achieved
in Ref.\,\cite{Hill}.  

Now, we demonstrate the existence of a considerable error in the above
scheme.
We solve Eq.\,(\ref{eq:flip_rot_eq}) without omitting
$H_{mix}^{i}$. 
After this calculation, we compare the exact solution with the approximate solution
(\ref{eq:rot_sol_approx}) by calculating the following quantity: 
\begin{equation}
F(\lvert\phi\rangle_{i})=|_{i}\langle\phi|e^{\frac{i}{\hbar}H^{i}_{rot}\,t_{X}}e^{-\frac{i}{\hbar}H^{i}_{d}\,t_{X}}|\phi\rangle_{i}|. \label{eq:overrap} 
\end{equation}
This is just the fidelity\,\cite{Nielsen} between the aimed state
$e^{-\frac{i}{\hbar}H^{i}_{d}\,t_{X}}|\phi\rangle_{i}$ obtained by an X
rotation and the physical state
$e^{-\frac{i}{\hbar}H^{i}_{rot}\,t_{X}}|\phi\rangle_{i}$ obtained by
Eq.\,(\ref{eq:flip_rot_eq}). 
If the value of $F(\lvert\phi\rangle_{i})$ is adequately close to one, the dynamics generated
only by $H^{i}_{d}$ is a good approximation for the genuine one; we can
achieve an X rotation by Hill and Goan's scheme. 
The exact solution can be represented by the eigenvalues and the
eigenvectors of Eq.\,(\ref{eq:sf_H_mrep}), which are given in
Appendix\,\ref{append:ana_Hrot}.
Adopting the typical value of the transverse magnetic field,
$B_{ac}=2.5\,{\rm mT}$, and choosing $\theta_{X}=\pi/4$, 
we calculate $F(\lvert\phi\rangle_{i})$ for three values of the HY:
$A/A_{0}=0.75,\,0.5,\,0.25$, and different two initial states. 
The X rotation for $\theta_{X}=\pi/4$ plays an essential role in the
Hadamard gate $U_{H}$:
$U_{H}=-i\,e^{i\pi\sigma_{z}/4}\,e^{i\pi\sigma_{x}/4}\,e^{i\pi\sigma_{z}/4}$. 
We show the results in Table \ref{tab:f_result}. 
\begin{table}
\caption{\label{tab:f_result}Calculated values of the fidelity (\ref{eq:overrap}) with $\theta_{X}=\pi/4$ and $B_{ac}=2.5\,\rm{mT}$.}
\begin{ruledtabular}
\begin{tabular}{ccc}
         &\multicolumn{2}{c}{$F(\lvert\phi\rangle_{i})$}\\
$A/A_{0}$&$\lvert\phi\rangle_{i}=|0\rangle_{i}$
&$\lvert\phi\rangle_{i}=(|0\rangle_{i}+|1\rangle_{i})/\sqrt{2}$ \\
\hline
0.75 & 0.72514 & 0.70458 \\
0.5  & 0.72458 & 0.70454  \\
0.25 & 0.73390 & 0.70526
\end{tabular}
\end{ruledtabular}
\end{table} 
They imply that the values of $F(\lvert\phi\rangle_{i})$ are
almost on the order of $10^{-1}$, regardless of the initial states and the
magnitude of the HY; the X rotation given by
Eq.\,(\ref{eq:vec_lab_approx}) involves a serious error, whose
probability, $1-\max_{\lvert\phi\rangle_{i}}[F(\lvert\phi\rangle_{i})]$,
is estimated at about $10^{-1}$.  
According to Ref.\,\cite{Hill}, the error probability in terms of the
fidelity (i.e.,
$1-\max_{\lvert\phi\rangle_{i}}[F(\lvert\phi\rangle_{i})]$) is estimated
about at $10^{-5}$ in the \CNOT gate. 
We will need to discuss the possibility of quantum error
correction\,\cite{Preskill}.
However, it is certain that the error in the X rotation in
Ref.\,\cite{Hill} is vastly larger than the one estimated there.

The presence of the error means that $H_{mix}^{i}$ in
Eq.\,(\ref{eq:sf_H_mrep}) shouldn't be regarded as a perturbation; the
bases where $H_{d}^{i}$ is a diagonal matrix are not suitable for the
analysis of the Schr\"odinger equation (\ref{eq:flip_rot_eq}). 
The above consideration is self--evident, investigating the time
evolution operator under the Lamour resonance condition (\ref{eq:Larmor}). 
The dynamics by $H_{d}^{i}$, which is relevant to the X rotation, is
characterized by the energy scale $\nu_{\theta}B_{ac}$. 
However, we usually find $\nu_{\theta}<\mu_{\theta}$. 
Consequently, the mixing effect by $H_{mix}^{i}$ becomes dominant. 
We also recognize the dominant effect of $H_{mix}^{i}$  from the
following viewpoint of estimating errors. 
If we simply apply the ordinal perturbation method to the calculation of
the eigenvectors of $H^{i}_{rot}$ in Eq.\,(\ref{eq:overrap}), we obtain
the following result for $\lvert\phi\rangle_{i}=\lvert 0\rangle_{i}$
under the condition (\ref{eq:Larmor}):  
\begin{equation*}
F(|0\rangle_{i}) 
= 
\sqrt{
\frac{1}{2}
+
\frac{1}{2}
\cos\left[
\frac{t_{X}}{\hbar}\left(
\tilde{\Delta}^{i}_{1}
-
\tilde{\Delta}^{i}_{0}
\right)
\right]}
+O\left(\frac{B_{ac}^{2}}{B^{2}}\right).
\end{equation*} 
Here, we regard $H_{d}^{i}$ as the free part in (\ref{eq:sf_H_mrep}). 
We define the difference between the lowest eigenvalue of $H_{d}^{i}$
and the one of $H_{rot}^{i}$ as $\tilde{\Delta}_{0}^{i}$.  
We also define the difference between the second lowest
eigenvalue of $H_{d}^{i}$ and the one of $H_{rot}^{i}$ as
$\tilde{\Delta}_{1}^{i}$.   
Indeed, we find  
$t_{X}(\tilde{\Delta}^{i}_{1}-\tilde{\Delta}^{i}_{0})/\hbar\approx 10^{-6}$
 for $A/A_{0}=0.5$ and $\theta_{X}=\pi/4$.
Then, we approximately get $F(|0\rangle_{i})\approx 1$ for 
$A/A_{0}=0.5$ and $\theta_{X}=\pi/4$; this result is clearly different
from the rigorous one. 

In summary, we have shown that the X rotation in Ref.\,\cite{Hill} includes a
considerable error, which is generated by $H_{mix}^{i}$.  
The mixing Hamiltonian $H_{mix}^{i}$
never appears under the assumption that the electron spin state is always
down, and this term should not be regarded as a perturbation for $H^{i}_{d}$. 
\subsection{The controlled--Z gates}
In Sec.\,\ref{subsec:spin_flip}, we explained that there is considerable
error in the X rotation carried out according to Hill and Goan's scheme.  
Generally, spin--flip operations play an important role in controlled
operations.
In Ref.\,\cite{Hill}, they are frequently used even in the controlled--Z
gates. 
Here, we show that the error in the controlled--Z gate mainly occurs in
the spin--flip operations.   

\begin{figure}[htbp]
\centering
\scalebox{0.5}[0.5]{\includegraphics{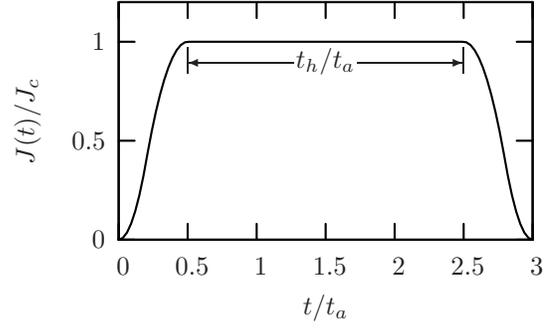}}
\caption{Profile of $J(t)$ ($\tau_{h}=t_{h}/t_{a}=2.0$ and $\tau^{\prime}=0.2$).} 
\label{fig:profile_J}
\end{figure}

First, by means of the adiabatic theorem, we calculate the time
evolution operator for two qubits during an interval $t_{C}$, 
\begin{equation}
U_{C}(t_{C})
=
T\left[\exp\left(
-\frac{i}{\hbar}\int_{0}^{t_{C}}H_{C}(t)\,dt\right)\right],
\label{eq:two_q_tev} 
\end{equation}
keeping $A_{1}(t)=A_{2}(t)=A_{0}$, turning off the transverse magnetic
field, and adiabatically varying $J(t)$. 
The total operation time of $J(t)$ is defined as $t_{C}$. 
In order to get an analytical expression of the operator, we take the
following profile of $J(t)$, introducing four parameters: The maximum
value of $J(t)$, $J_{c}$, the period $t_{h}$ when $J(t)$ keeps $J_{c}$,
the former and the latter periods when $J(t)$ varies adiabatically,
$t_{a}$ (i.e., $t_{C}=t_{a}+t_{h}$), and the smoothness of profile
touching to the constant line, $\tau^{\prime}$. 
The assumed time dependence of $J(t)$ is as follows: 
\begin{equation}
J(t)=
\left\{
\begin{array}{ll}
J_{c}f_{J}(\tau) & \left(0\le\tau\le\frac{1}{2}\right)\\
J_{c} & \left(\frac{1}{2}\le\tau\le\tau_{C}-\frac{1}{2}\right)  \\
J_{c}f_{J}(\tau_{C}-\tau) & \left(\tau_{C}-\frac{1}{2}\le\tau\le \tau_{C}\right)
\end{array}
\right., \label{eq:J_profile}
\end{equation}
where $\tau$ is a scaled time $\tau=t/t_{a}$, $\tau_{C}=t_{C}/t_{a}$,
and $\tau_{h}=t_{h}/t_{a}$ (see Fig.\,\ref{fig:profile_J}). 
The function $f_{J}(\tau)$ in Eq.\,(\ref{eq:J_profile}) is defined as, 
\begin{equation*}
f_{J}(\tau)
=
\Theta(\tau^{\prime}-\tau)\frac{2}{\tau^{\prime}}\tau^{2}
+
\Theta(\tau-\tau^{\prime})\left[
1-\frac{(1-2\tau)^{2}}{1-2\tau^{\prime}}
\right].
\end{equation*}
The initial state and final state for two qubits will be spanned by $\lvert v_{1}\rangle$,
$\lvert v_{+}\rangle$, $\lvert v_{-}\rangle$, and $\lvert v_{4}\rangle$,
because $J(0)=J(t_{C})=0$ and $A_{1}(t)=A_{2}(t)=0$ ($0\le t\le t_{C}$). 
Then, the time--evolution operator (\ref{eq:two_q_tev}) is given by  
\begin{equation}
U_{C}(t_{C})
=
\sum_{k=1,\,+,\,-,\,4}e^{-i(\alpha_{k}+\beta_{k})}\,
\lvert v_{k}\rangle\langle v_{k}\rvert
+
R_{C}. \label{eq:two_ad_t_ev}
\end{equation}
The phase $\alpha_{k}$ and $\beta_{k}$ in Eq.\,(\ref{eq:two_ad_t_ev})
are related to the processes in which $J(t)$ varies (i.e., $0\le \tau\le 1/2$ and
$\tau_{C}-1/2\le\tau\le\tau_{C}$) and in which $J(t)=J_{c}$ (i.e.,
$1/2\le\tau\le\tau_{C}-1/2$), respectively.  
The analytical expressions for them are shown in Appendix
\ref{append:ana_eq}. 
The last term in Eq.\,(\ref{eq:two_ad_t_ev}), like the third term in
Eq.\,(\ref{eq:s_ad_t_ev}), represents the deviation from the adiabatic
approximation and mixes the states relevant to quantum computation with
the irrelevant states. 
This gives a dominant contribution near the level crossing region around
$J_{c}=\mu_{B}B/2$ (see, Fig.\,\ref{fig:c_elv}), where we have to apply
carefully the adiabatic approximation.   
In this paper, we investigate Eq.\,(\ref{eq:two_q_tev}) in the regime
where the level crossing is not included; we can
safely neglect $R_{C}$ in Eq.\,(\ref{eq:two_ad_t_ev}). 
Actually, we confirm that the adiabatic approximation works very well,
solving numerically the time--dependent Schr\"odinger equation.  

Secondly, we explain how to connect $U_{C}(t_{C})$ with a
controlled--Z gate.  
Hereafter, we assume that the spin--flip operations are realized by using
Eq.\,(\ref{eq:vec_lab_approx}).  
First of all, we rewrite Eq.\,(\ref{eq:two_ad_t_ev}) as 
$U_{C}(t_{C})= U_{ad}U_{st}$, 
where 
$U_{ad}=\sum_{k}e^{-i\alpha_{k}}\lvert v_{k}\rangle\langle v_{k}\rvert$ 
and 
$U_{st}=\sum_{k}e^{-i\beta_{k}}\lvert v_{k}\rangle\langle v_{k}\rvert$. 
Next, we investigate $U_{st}$ in more detail. 
Introducing 
$\openone_{i}\equiv|0\rangle_{L,i}\langle 0|+|1\rangle_{L,i}\langle 1|$, 
$\sigma_{i\,z}\equiv|0\rangle_{L,i}\langle 0|-|1\rangle_{L,i}\langle 1|$, 
$\sigma_{i\,x}\equiv|0\rangle_{L,i}\langle 1|+|1\rangle_{L,i}\langle 0|$, and 
$\sigma_{i\,y}\equiv-i(|0\rangle_{L,i}\langle 1|-|1\rangle_{L,i}\langle 0|)$, 
we find that $U_{st}$ is given by  
\begin{equation*}
U_{st}
=
V\,e^{-i\delta_{C}(\sigma_{1\,x}\otimes\sigma_{2\,x}+\sigma_{1\,y}\otimes\sigma_{2\,y})}
e^{-i\delta_{C}^{\prime}\sigma_{1\,z}\otimes\sigma_{2\,z}},
\end{equation*}
where 
$V=e^{-ic}\,e^{-i\delta_{s}(\sigma_{1\,z}+\sigma_{2\,z})}$,
$c=(\beta_{1}+\beta_{+}+\beta_{-}+\beta_{4})/4$, 
$\delta_{s}=(\beta_{1}-\beta_{4})/4$, 
$\delta_{C}=(\beta_{+}-\beta_{-})/4$, and 
$\delta_{C}^{\prime}=(\beta_{1}-\beta_{+}-\beta_{-}+\beta_{4})/4$. 
Note that this operator contains spin--flip operations. 
Here, we define the unitary operator $W$ as 
\begin{eqnarray*}
W
&=&
(U_{H}\otimes U_{H})V^{\dagger}U_{st}(U_{H}\otimes U_{H}) \nonumber \\
&=&
e^{-i\delta_{C}(\sigma_{1\,z}\otimes\sigma_{2\,z}+\sigma_{1\,y}\otimes\sigma_{2\,y})}
e^{-i\delta^{\prime}_{C}\sigma_{1\,x}\otimes \sigma_{2\,x}},
\end{eqnarray*}
where $U_{H}$ is the Hadamard gate.  
Then, according to the idea shown in Ref.\,\cite{Hill}, we find that the
controlled--Z gate $U_{cz}(\theta_{cz})$ of the angle $\theta_{cz}$ is
achieved as follows:  
\begin{eqnarray}
U_{cz}(\theta_{cz})
&=&
e^{-i\frac{\theta_{cz}}{2}\sigma_{1\,z}\otimes\sigma_{2\,z}}(\openone_{1}\otimes
e^{i\frac{\theta_{cz}}{2}\sigma_{2\,z}}) \nonumber \\
&=&
W(\openone_{1}\otimes\sigma_{2\,z})W(\openone_{1}\otimes\sigma_{2\,z})
\nonumber \\
&&
\quad\quad\quad\quad\quad\quad\quad
\times
(\openone_{1}\otimes e^{i\frac{\theta_{cz}}{2}\sigma_{2\,z}}).
\label{eq:rel_st_cz} 
\end{eqnarray} 
Note that the several basic identities
$\sigma_{i\,z}\sigma_{i\,x}\sigma_{i\,z}=-\sigma_{i\,x}$,
$\sigma_{i\,z}\sigma_{i\,y}\sigma_{i\,z}=-\sigma_{i\,y}$, and 
$[\sigma_{1\,a}\otimes\sigma_{2\,a},\,\sigma_{1\,b}\otimes\sigma_{2\,b}]=0$ 
($a,\,b=x,\,y,\,z$) 
play an essential role in the derivation of Eq.\,(\ref{eq:rel_st_cz}). 
The angle $\theta_{cz}$ is given by
$\theta_{cz}=4\delta_{C}=\beta_{+}-\beta_{-}$, from which $t_{h}$ is given by,   
\begin{equation}
t_{h}=\frac{\theta_{cz}\hbar}{2J_{c}+\sqrt{(\epsilon-2J_{c})^{2}+4A_{0}^{2}}-\sqrt{\epsilon^{2}+4A_{0}^{2}}}.\label{eq:theta_cz}
\end{equation}
Here, let us recall that the physical time--evolution is given by
$U_{C}(t_{C})$. 
The unitary operator $W$ in Eq.\,(\ref{eq:rel_st_cz}) is rewritten in
the term of $U_{C}(t_{C})$ as follows: 
\begin{equation}
W
=
(U_{H}\otimes U_{H})V^{\dagger}U^{\dagger}_{ad}U_{C}(t_{C})(U_{H}\otimes U_{H}). 
\label{eq:W_U_C}
\end{equation}
The unitary operator $V$ is composed of several Z rotations.

Now that we find the connection between $U_{C}(t_{C})$ and the
controlled--Z gate, we should discuss the effect of $U_{ad}$. 
Generally, the unitary operator $U_{ad}$ should include two qubit
operations. 
We would like to represent $U_{ad}$ as single qubit operations, in
particular, Z rotations, because they have been constructed in
Sec.\,\ref{subsec:PS}, and their errors are very small. 
We find that $U_{ad}$ is equal to the single qubit operation, 
$\sigma_{1\,z}\otimes\sigma_{2\,z}$, 
if the following conditions are fulfilled:
$\alpha_{+}-\alpha_{1}=\pi+2m_{+}\pi$, 
$\alpha_{-}-\alpha_{1}=\pi+2m_{-}\pi$, 
and
$\alpha_{4}-\alpha_{1}=2m_{4}\pi$ 
($m_{+},\,m_{-},\,m_{4}\in\mathbb{Z}$ and $m_{4}\neq 0$). 
These three conditions are written by  
\begin{widetext}
\begin{eqnarray}
&&
\frac{
-\sqrt{\epsilon^{2}+4A_{0}^{2}}-\epsilon+2g_{n}\mu_{n}B
+2\int_{0}^{\frac{1}{2}}\left[J(\tau)-E_{1}^{(0,\,+)}(\tau)\right]\,d\tau 
}{
-2\epsilon+4g_{n}\mu_{n}B+2A_{0}
+2\int_{0}^{\frac{1}{2}}\left[J(\tau)-E_{1}^{(0,\,+)}(\tau)\right]\,d\tau 
} 
=
\frac{1+2m_{+}}{2m_{4}}, \label{eq:alpha_cond_1}\\
&&
\frac{
-\epsilon+2g_{n}\mu_{n}B
-
2\int_{0}^{\frac{1}{2}}
\left\{
\sqrt{[\epsilon-2J(\tau)]^{2}+4A_{0}^{2}}+J(\tau)+E_{1}^{(0,\,+)}(\tau)\right\}\,d\tau 
}{
-2\epsilon+4g_{n}\mu_{n}B+2A_{0}
+2\int_{0}^{\frac{1}{2}}\left[J(\tau)-E_{1}^{(0,\,+)}(\tau)\right]\,d\tau 
} 
=
\frac{1+2m_{-}}{2m_{4}}, \label{eq:alpha_cond_2}\\
&&
t_{a}
=
\frac{2m_{4}\pi\hbar}{
-2\epsilon+4g_{n}\mu_{n}B+2A_{0}
+2\int_{0}^{\frac{1}{2}}\left[J(\tau)-E_{1}^{(0,\,+)}(\tau)\right]\,d\tau 
}.  \label{eq:alpha_cond_3}
\end{eqnarray}
\end{widetext}
The analytical expression for $E_{1}^{(0\,,+)}(\tau)$ is shown in Appendix
\ref{append:ana_eq}. 
We find that Eqs.\,(\ref{eq:alpha_cond_1}) and (\ref{eq:alpha_cond_2})
include two unknown parameters $J_{c}$ and $\tau^{\prime}$, besides the
free parameters $m_{+}$, $m_{-}$, and $m_{4}$. 
We should choose suitable values for the free parameters that bring about
a satisfactory solution. 
First, we numerically find that both the right--hand sides of Eqs.\,(\ref{eq:alpha_cond_1})
and (\ref{eq:alpha_cond_2}) are close to $1/2$ for $0<J_{c}<\epsilon/2$
and $0<\tau^{\prime}<1/2$.  
Thus, we choose $m_{+}=m_{-}=0$ and $m_{4}=1$. 
After this consideration, we numerically solve the simultaneous systems of equation composed of
Eqs.\,(\ref{eq:alpha_cond_1}) and (\ref{eq:alpha_cond_2}) by the
Newton--Raphson method. 
Then, we obtain $t_{a}$ by substituting the value of
$m_{4}$, $J_{c}$, and $\tau^{\prime}$ for Eq.\,(\ref{eq:alpha_cond_3}). 
Finally, the remaining parameter $t_{h}$ is determined by
Eq.\,(\ref{eq:theta_cz}). 
 
We show the typical solutions for $\theta_{cz}=\pi$ in
Table~\ref{tab:cz_result}. 
The case of $\theta_{cz}=\pi$ is very important, because a {\scriptsize
CNOT} gate is essentially composed of $U_{cz}(\pi)$ and Hadamard gates.  
We find that if we choose a large value for $J_{c}$, the corresponding
values of $\tau^{\prime}$ and $t_{h}$ decrease; the operation time of the
controlled--Z gate becomes short. 
On the other hand, the value of $t_{a}$ for each $J_{c}$ is almost the 
same and greatly shorter than $t_{h}$; it suggests that we can
safely regard the adiabatic controlling part of the profile of $J(t)$ (i.e.,
$0\le\tau\le1/2$ and $\tau_{C}-1/2\le\tau\le\tau_{C}$) as an
instantaneous process within the total controlling one. 

Let us summarize the above discussion. 
First, we obtained Eq.\,(\ref{eq:rel_st_cz}), which represents the
relation between the logical controlled--Z gate and the physical time
evolution, assuming the validity of the adiabatic approximation and the
realization of the spin--flip operations by Eq.\,(\ref{eq:rel_st_cz}). 
The result is essentially equivalent to that of Ref.\,\cite{Hill}, except
for the presence of $U_{ad}$. 
Here, we have shown that $U_{ad}$ is equivalent to the single
qubit operation, $\sigma_{1\,z}\otimes\sigma_{2\,z}$, if we choose a 
suitable set of the parameters $(J_{c},\,\tau^{\prime},\,t_{a},\,t_{h})$.  
We find that one of the two assumptions, the adiabatic approximation, is
valid, because we choose a value of $J_{c}$ which is far from the
level crossing point. 
Unfortunately, the spin--flip operations by Eq.\,(\ref{eq:rel_st_cz})
don't quite work. 
The main origin of the error in the controlled--Z gates lies in the
part relevant to the spin--flip operations (e.g., the Hadamard gate). 
\begin{table}
\caption{\label{tab:cz_result}Calculated values of $J_{c}$, $\tau^{\prime}$, $t_{a}$, and $t_{h}$ with $\theta_{cz}=\pi$, $m_{+}=0$, $m_{-}=0$, and $m_{4}=1$. }
\begin{ruledtabular}
\begin{tabular}{cccc}
$J_{c}/\epsilon$ & $\tau^{\prime}$ & $t_{a}$ (${\rm n}s$) & $t_{h}$ ($\mu s$)\\
\hline
0.1003 & 0.1085 & 5.391 & 33.1 \\
0.1988 & 0.0916 & 5.391 & 12.59 \\
0.01 & 0.2203   & 5.392 & 40.44 
\end{tabular}
\end{ruledtabular}
\end{table}
\section{Summary}\label{sec:summary}
We have re--investigated the operating schemes of quantum gates associated
with universal quantum gates (i.e., Z rotations, X rotations, and controlled--Z
gates), for Kane's model in detail.
We chose the suitable computational bases out of the eigenstates of
$H^{i}(t)$ with fixed time.   
Our choice is different from those in Refs.\,\cite{Wellard, Fowler, Hill}, in which
the electron spin state is assumed to be always down. 
The bases in the latter case are fragile for the time--evolution due to
$H^{i}(t)$. 
The phase--shift operations (i.e., Z rotations) for the $i$ th qubit are
shown to be constructed with an extremely low error probability,
adiabatically varying the value of $A_{i}(t)$ with the fixed other
parameters: $J(t)=0$, $A_{j}(t)=A_{0}$, and $B_{ac}=0$.
The most important result is that a considerable error for the X
rotation in Ref.\,\cite{Hill} exists. 
The physical origin of such an error is the mixing effect between the
computational bases and the irrelevant states by $H_{mix}^{i}$, which is
neglected in Ref.\,\cite{Hill}. 
This issue emerged from discussion of Eq.\,(\ref{eq:flip_rot_eq}), not by 
perturbation but by a rigorous analysis.   
Finally, the controlled operations in Ref.\,\cite{Hill} will not quite work,
because the spin--flip operations are used in them. 
However, it is meaningful to check whether they contain other errors. 
In particular, we have investigated the controlled--Z gate, as it is an
essential part of the \CNOT gate.  
We have shown that the main origin of the error in the controlled--Z
gate lies in the parts relevant to the spin--flip operations. 
Accordingly, we will have to discuss recovery of the errors caused by the
spin--flip operations.

It is necessary for quantum computation to control quantum systems. 
In doing this, it is inevitable that some errors occur in the controlling
processes, even if we do them carefully. 
Therefore, it is crucial to make clear the physical origin of the errors
and to investigate methods of correcting them.
Kane's model is an attractive proposal for quantum computers. 
In this paper, we have presented the robust quantum computational bases
for $H^{i}(t)$ time--evolution, but they do not quite work for
the spin--flip operations.
We will need to find more suitable computational bases for the
operations of quantum gates. 
In addition, we will have to take account of the modified versions of 
Kane's model\,\cite{Skinner, Hill2004}.  
\begin{acknowledgments}
The authors acknowledge H. Nakazato, K. Yuasa, T. Watanabe, and
 T. Shinada for valuable discussions.    
This work is supported partially by a Grant--in--Aid for the COE Research at
Waseda University and that for Priority Area B (No.\,763), MEXT, by
a Grant for the 21st Century COE Program at Waseda University, and by a
Waseda University Grant for Special Research Projects (No.\,2004B--872).
The authors thank the Yukawa Institute for Theoretical Physics at Kyoto University. 
Discussions during the YITP workshop YITP--W--04--14 on ``Chaos and
 Nonlinear Dynamics in Quantum-Mechanical and Macroscopic Systems'' were
 useful to the completion of this work. 
\end{acknowledgments}
\appendix
\section{The instantaneous eigenvalues of the two qubit Hamiltonian}\label{append:ana_eq}
We show the explicit expressions for the instantaneous eigenvalues of
$H_{C}(t)$ at the time $t$ if $A_{1}(t)=A_{2}(t)(\equiv A(t))$. 
The Hamiltonian $H_{C}(t)$ is a $16\times 16$ matrix and is decomposed
into several submatrices by the system's symmetry (see
Eq.\,(\ref{eq:c_h_dcmp})). 
Hereafter, we abbreviate the argument $t$.  
We only write the eigenvalues relevant to the computation bases:
\begin{eqnarray}
E^{(0,1)}_{1} 
&=& 
-A-\frac{1}{2}\sqrt{\xi-c_{1}} \nonumber \\
&&
\quad\quad
-\frac{1}{2}\sqrt{-\xi-c_{1}+\frac{2c_{2}}{\sqrt{\xi-c_{1}}}}, 
\label{eq:E_0_1_1}\\ 
E^{(0,1)}_{2} 
&=& 
-A-\frac{1}{2}\sqrt{\xi-c_{1}} \nonumber \\
&&
\quad\quad
+ \frac{1}{2}\sqrt{-\xi-c_{1}+\frac{2c_{2}}{\sqrt{\xi-c_{1}}}}, 
\label{eq:E_0_1_2}\\
E^{(-1,\,-1)}_{1} 
&=&
-J-\epsilon+2g_{n}\mu_{n}B \nonumber \\
&&
\quad\quad
-\sqrt{(\epsilon-2J)^{2}+4A^{2}}, 
\label{eq:E_-1_-1_1}\\
E^{(-1,\,-1)}_{2} 
&=&
-J-\epsilon+2g_{n}\mu_{n}B \nonumber \\
&&
\quad\quad
+ \sqrt{(\epsilon-2J)^{2}+4A^{2}}, 
\label{eq:E_-1_-1_2}\\
E^{(-1,\,1)}_{1}
&=&
J-\epsilon+2g_{n}\mu_{n}B-\sqrt{\epsilon^{2}+4A^{2}}, 
\label{eq:E_-1_1_1}\\ 
E^{(-2,\,1)}_{1}
&=&
J-2\epsilon+4g_{n}\mu_{n}+2A. 
\label{eq:E_-2_1_1}
\end{eqnarray} 
We define $\xi$ in Eqs.\,(\ref{eq:E_0_1_1}) and (\ref{eq:E_0_1_2}) as
a real root of the following cubic equation: 
\begin{equation}
z^{3} -c_{1}z^{2} -4c_{3}z + 4c_{3}c_{1} - c_{2}^{2} = 0. \label{eq:3rd_eq_a} 
\end{equation}
Taking account of Cardano's formula\,\cite{Tignol}, we easily obtain a
real root of Eq.\,(\ref{eq:3rd_eq_a}). 
The coefficients in Eqs.\,(\ref{eq:E_0_1_1}), (\ref{eq:E_0_1_2}), and
(\ref{eq:3rd_eq_a}), $c_{1}$, $c_{2}$, and $c_{3}$, are given by,  
\begin{eqnarray*}
c_{1} 
&=&
-6J^{2}+4JA-22A^{2}-4\epsilon^{2}, \\
c_{2}
&=&
8J^{3}-8AJ^{2}+8(A^{2}-\epsilon^{2})J+8\epsilon^{2}A+24A^{3}, \\
c_{3}
&=&
-3J^{4} + 4AJ^{3}+(14A^{2}+12\epsilon^{2})J^{2} \\
&&
\quad\quad
+(-60A^{3}+8\epsilon^{2}A)J + 12\epsilon^{2}A^{2}+45A^{4}.
\end{eqnarray*}
Using Eqs.\,(\ref{eq:J_profile}) and
(\ref{eq:E_0_1_1})--(\ref{eq:E_-2_1_1}), the phases $\alpha_{k}$ and
$\beta_{k}$ in Eq.(\ref{eq:two_ad_t_ev}) are given 
as follows: 
\begin{eqnarray*}
\alpha_{1} &=& \frac{2t_{a}}{\hbar} \int_{0}^{1/2}E^{(0,\,1)}_{1}(J(\tau))d\tau, \\
\alpha_{\pm} &=& \frac{2t_{a}}{\hbar} \int_{0}^{1/2}E^{(-1,\,\pm 1)}_{1}(J(\tau)) d\tau, \\
\alpha_{4} &=& \frac{2t_{a}}{\hbar} \int_{0}^{1/2}E^{(-2,\,1)}_{1}(J(\tau))d\tau, \\
\beta_{1} &=& \frac{t_{h}}{\hbar} E^{(0,\,1)}_{1} (J_{c}), \\
\beta_{\pm} &=& \frac{t_{h}}{\hbar} E^{(-1,\,\pm 1)}_{1} (J_{c}), \\
\beta_{4} &=& \frac{t_{h}}{\hbar} E^{(-2,\,1)}_{1} (J_{c}).
\end{eqnarray*}
\section{The solution of the eigenvalue problem for the single qubit
 Hamiltonian in the rotating frame}\label{append:ana_Hrot}
Let us consider the solution for the eigenvalue problem,
$H^{i}_{rot}\lvert w_{k}\rangle_{i} = \hbar\Omega^{i}_{k}\lvert w_{k}\rangle_{i}$ 
($k=0,\,1,\,2,\,3$). 
The eigenvalues are given by 
\begin{eqnarray}
\hbar\Omega^{i}_{0}
&=&
-\frac{1}{2}\sqrt{\eta-d_{1}} \nonumber\\
&&\quad\quad
-\frac{1}{2}\sqrt{-\eta-d_{1}+\frac{2d_{2}}{\sqrt{\eta-d_{2}}}},\label{eq:Omega_1}\\
\hbar\Omega^{i}_{1}
&=&
-\frac{1}{2}\sqrt{\eta-d_{1}}\nonumber \\
&&\quad\quad
+\frac{1}{2}\sqrt{-\eta-d_{1}+\frac{2d_{2}}{\sqrt{\eta-d_{2}}}},\label{eq:Omega_2}\\
\hbar\Omega^{i}_{2}
&=&
\frac{1}{2}\sqrt{\eta-d_{1}} \nonumber \\
&&\quad\quad
-\frac{1}{2}\sqrt{-\eta-d_{1}-\frac{2d_{2}}{\sqrt{\eta-d_{2}}}},\label{eq:Omega_3}\\
\hbar\Omega^{i}_{3}
&=&
\frac{1}{2}\sqrt{\eta-d_{1}} \nonumber \\
&&\quad\quad
+\frac{1}{2}\sqrt{-\eta-d_{1}-\frac{2d_{2}}{\sqrt{\eta-d_{2}}}} \label{eq:Omega_4}. 
\end{eqnarray} 
We define $\eta$ in Eqs.\,(\ref{eq:Omega_1})--(\ref{eq:Omega_4}) as a
real root of the following cubic equation:  
\begin{equation}
z^{3}-d_{1}z^{2}-4d_{3}z+4d_{1}d_{3}-d_{2}^{2}=0. \label{eq:3rd_eq_b} 
\end{equation}
We easily find a real root of Eq.\,(\ref{eq:3rd_eq_b}) through Cardano's
formula\,\cite{Tignol}, as in Appendix \ref{append:ana_eq}. 
The coefficients, $d_{1}$, $d_{2}$, and $d_{3}$ in
Eqs.\,(\ref{eq:Omega_1})--(\ref{eq:3rd_eq_b}), are given by  
\begin{eqnarray*}
d_{1} 
&=&
E^{i}_{0}E^{i}_{2}-(E^{i}_{0}+E^{i}_{2})^{2}
+(E^{i}_{1}-\hbar\omega_{ac})(E^{i}_{3}+\hbar\omega_{ac}) \\
&&\quad\quad
-2(\mu_{B}B_{ac})^{2}-2(g_{n}\mu_{n}B_{ac})^{2}, \\
d_{2}
&=&
(E^{i}_{0}+E^{i}_{2})\left[E^{i}_{0}E^{i}_{2}-(E^{i}_{1}-\hbar\omega_{ac})(E^{i}_{3}+\hbar\omega_{ac})\right] \\
&&\quad\quad
+ 4g_{n}\mu_{n}\mu_{B}B^{2}_{ac}A, \\
d_{3}
&=&
E^{i}_{0}E^{i}_{2}(E^{i}_{1}-\hbar\omega_{ac})(E^{i}_{3}+\hbar\omega_{ac})
+(\mu_{B}^{2}-g_{n}^{2}\mu_{n}^{2})^{2}B_{ac}^{2} \\
&&\quad\quad
-(E^{i}_{1}-\hbar\omega_{ac})(E^{i}_{0}\nu_{-\theta}^{2}+E^{i}_{2}\mu_{-\theta}^{2})B_{ac}^{2}\\
&&\quad\quad
-(E^{i}_{3}+\hbar\omega_{ac})(E^{i}_{0}\mu_{\theta}^{2}+E^{i}_{2}\nu_{\theta}^{2})B_{ac}^{2}.
\end{eqnarray*}
The eigenvector $\lvert w_{k}\rangle_{i}$ associated with the eigenvalue
$\hbar\Omega^{i}_{k}$ is given by 
\begin{widetext}
\begin{eqnarray*}
\lvert w_{k}\rangle_{i}
&=&
\mathcal{N}\bigg[\left(
-\frac{\nu_{-\theta}}{\mu_{-\theta}}\frac{E^{i}_{0}-\hbar\Omega^{i}_{k}}{\nu_{\theta}B_{ac}} 
-
\frac{E^{i}_{2}-\hbar\Omega^{i}_{k}}{\mu_{\theta}B_{ac}}
\right)\lvert u_{0}(A)\rangle_{i} \\
&&
+
\left(
\frac{\mu_{-\theta}}{\nu_{\theta}}-\frac{\nu_{-\theta}}{\mu_{\theta}}-\frac{E^{i}_{3}+\hbar\omega_{ac}-\hbar\Omega^{i}_{k}}{\mu_{-\theta}B_{ac}}\frac{E^{i}_{0}-\hbar\Omega^{i}_{k}}{\nu_{\theta}B_{ac}}\right)
\lvert u_{1}(A)\rangle_{i}\\
&&
+
\left(
\frac{E^{i}_{3}+\hbar\Omega_{ac}-\hbar\Omega^{i}_{k}}{\mu_{-\theta}B_{ac}}
\frac{E^{i}_{2}-\hbar\Omega^{i}_{k}}{\mu_{\theta}B_{ac}}
+\frac{\nu_{-\theta}}{\nu_{\theta}}
-\frac{\nu_{-\theta}^{2}}{\mu_{\theta}\mu_{-\theta}}
\right)\lvert u_{2}(A)\rangle_{i} \\
&&
+
\left(
\frac{E^{i}_{3}+\hbar\omega_{ac}-\hbar\Omega^{i}_{k}}{\mu_{\theta}B_{ac}}
\frac{E^{i}_{2}-\hbar\Omega^{i}_{k}}{\mu_{-\theta}B_{ac}}
\frac{E^{i}_{0}-\hbar\Omega^{i}_{k}}{\nu_{\theta}}
-\frac{\nu_{-\theta}^{2}}{\mu_{\theta}\mu_{-\theta}}
\frac{E^{i}_{0}-\hbar\Omega^{i}_{k}}{\nu_{\theta}}
-\frac{\mu_{-\theta}}{\mu_{\theta}}
\frac{E^{i}_{2}-\hbar\Omega^{i}_{k}}{\nu_{\theta}}
\right)\lvert u_{3}(A)\rangle_{i}\bigg], 
\end{eqnarray*}
\end{widetext}
where $\mathcal{N}$ is a normalization factor.

\end{document}